\documentclass[twocolumn]{revtex4}%
\usepackage{amsfonts}
\usepackage{amsmath}
\usepackage{amssymb}
\usepackage{graphicx}%
\providecommand{\U}[1]{\protect\rule{.1in}{.1in}}
\providecommand{\U}[1]{\protect\rule{.1in}{.1in}}

\providecommand{\U}[1]{\protect\rule{.1in}{.1in}}
\providecommand{\U}[1]{\protect\rule{.1in}{.1in}}
\begin{document}
\title{Hermite-Halphen-Bloch solution of two-gap Lam\'e equation}
\author{Jun-ichiro Kishine}
\affiliation{Division of Natural and Environmental Sciences, The Open University of Japan,
Chiba, 261-8586, Japan}
\author{A. S. Ovchinnikov}
\affiliation{Institute of Natural Sciences, Ural Federal University, Ekaterinburg, 620083, Russia}
\date{\today }

\begin{abstract}
We construct the Hermite--Halphen--Bloch solution for the two-gap ($n=2$)
Lam\'e equation of Jacobian form and give closed formulae to calculate the
energy band dispersion relation.

\end{abstract}
\maketitle

The Jacobian form of the Lam\'{e} equation,\cite{Whittaker-Watson}%
\begin{align}
\left[  -\frac{d^{2}}{dx^{2}}+n(n+1)\kappa^{2}\mathrm{sn}^{2}x\right]
\psi(x)=\varepsilon\psi(x). \label{Lame_equation}%
\end{align}
has wide application in physics, where the Jacobi elliptic functions
$\mathrm{sn}x=\mathrm{sn}(x,\kappa)$, $\mathrm{cn}x=\mathrm{cn(}x,\kappa)$,
and $\mathrm{dn}x=\mathrm{dn(}x,\kappa)$ are doubly-periodic functions in the
complex plane, with modulus $\kappa$ ($0\leq\kappa\leq1$). The Lam\'{e}
equation appears in a wide range of
physics\cite{Alhassid1983,Liang1992,Musevic1994,Sutcliffe1996,Dunne1998,Sakamoto2000,Li2000,Bena2006}
. For example, the Gaussian (one-loop) fluctuations in the one-dimensional
sine-Gordon model\cite{Sutherland1973} and the $\varphi^{4}$%
-model\cite{Pawellek2009} obey respectively 1-gap and 2-gap Lam\'{e} equations.

The solutions of Eq. (\ref{Lame_equation}) for positive integer $n$ are given
by
\begin{equation}
\psi(x)=\prod_{j=1}^{n}\left[  \frac{\mathrm{H}(x+\alpha_{j})}{\Theta
(x)}e^{-xZ(\alpha_{j})}\right]  , \label{LameSolution}%
\end{equation}
which is referred to as a Hermite--Halphen
solution\cite{Whittaker-Watson,Maier2008}. Here $\Theta,$ $\mathrm{H}$ and $Z$
are Jacobi's Theta, Eta and Zeta functions, with periods $4K,2K,2K,$
respectively, where $K=K\left(  \kappa\right)  $ is the complete elliptic
integral of the first kind. The complex parameters $\alpha_{1},\alpha
_{2},...\alpha_{n}$ are determined by the constraints
equations,\cite{Whittaker-Watson}
\begin{equation}
\varepsilon=\sum_{j=1}^{n}\frac{1}{\mathrm{sn}^{2}\alpha_{j}}-\left[
\sum_{j=1}^{n} \frac{\mathrm{cn}\alpha_{j}\mathrm{dn}\alpha_{j}}%
{\mathrm{sn}\alpha_{j}} \right]  ^{2}, \label{Condition1}%
\end{equation}%
\begin{equation}
\sum_{j=1}^{n} \frac{\mathrm{sn}\alpha_{j}\mathrm{cn}\alpha_{j}\mathrm{dn}%
\alpha_{j}+\mathrm{sn}\alpha_{p}\mathrm{cn}\alpha_{p}\mathrm{dn}\alpha_{p}%
}{\mathrm{sn}^{2}\alpha_{j}-\mathrm{sn}^{2}\alpha_{p}}=0\quad(j\neq p).
\label{Condition2}%
\end{equation}

When $x$ is restricted to real axis, the equation (\ref{Lame_equation}) is
regarded as a Schr\"{o}dinger equation for a particle moving in a
one-dimensional potential, $V(x)=\kappa^{2}n(n+1)\mathrm{sn}^{2}x,$ which is
bounded and periodic\ with its period being $2K$. Therefore, Eq.
(\ref{Lame_equation}) is a kind of Hill's equation\cite{Arscott64}. According
to standard results on Hill's equation, imposing a quasi-periodic boundary
condition
\begin{equation}
\psi(x+2K)=e^{-i(2K)k}\psi(x)\equiv\xi\psi(x), \label{Bloch}%
\end{equation}
with a real parameter $k$ being fixed and a Floquet multiplier satisfying
$|\xi|=1$, defines a self-adjoint boundary value problem and there exists the
Bloch-wave solution with crystal momentum $k$. For a positive integer $n$,
$V(x)$ is called the $n$-gap Lam\'{e} potential, because the Bloch spectrum
consists of the $n+1$ bands $\varepsilon_{1}\leq\varepsilon\leq\varepsilon
_{2},$ $\varepsilon_{3}\leq\varepsilon\leq\varepsilon_{4},\dots,\varepsilon
_{2n+1}\leq\varepsilon<\infty$\cite{Ince40}, i.e., there are $n$ finite bands
followed by a continuum band, separated by $n$ forbidden lacuna. The
Bloch-wave functions at the $2n+1$ band edges $\varepsilon_{1},\dots
,\varepsilon_{2n+1}$ are periodic and anti-periodic depending of the Floquet
multiplier being $\xi=+1$ or $\xi=-1$ and are represented in a form of
polynomials (so called Lam\'{e} polynomials)\ in $\mathrm{sn}$, $\mathrm{cn}%
x$, and $\mathrm{dn}$ functions. Comparing the general solution
(\ref{LameSolution}) and the Bloch form (\ref{Bloch}), we find
\begin{equation}
k=-i\sum_{j=1}^{n}Z(\alpha_{j})+\frac{n\pi}{2K}. \label{wavenumber}%
\end{equation}
The expression for the momentum in the Weierstrass form of the Lam\'{e}
equation is given in the Appendix A. The allowed energy bands correspond to a
real value of the wavenumber $k$, i.e., to the condition
\begin{equation}
\operatorname{Re}\left[  \sum_{j=1}^{n}Z(\alpha_{j},k)\right]  =0.
\label{condition_for_wave_number}%
\end{equation}
Now, a full set of equations (\ref{LameSolution}), (\ref{Condition1}),
(\ref{Condition2}), (\ref{wavenumber}), and (\ref{condition_for_wave_number})
give the Bloch\ band dispersion relations, i.e., $\varepsilon=\varepsilon(k)$.

The physical origin of the $n$-gap band structure is understood by observing,
\begin{equation}
\kappa^{2}\mathrm{\mathrm{sn}}^{2}x=-\left(  {\frac{\pi}{2\bar{K}}}\right)
^{2}\sum_{\ell=-\infty}^{\infty}\mathrm{sech\,}^{2}\left[  \frac{\pi}{2\bar
{K}}\left(  x-2K\ell\right)  \right]  +\frac{\bar{E}}{\bar{K}}.
\label{decompositionfourmula}%
\end{equation}
[derivation of this formula is given in Appendix B]. This relation indicates
that the Lam\'{e} potential consists of a periodic array of the the modified
P\"{o}schl-Teller potential\cite{Flugge} centered at $x=2K\ell$,
\begin{equation}
U_{\ell}(x)=-n(n+1)\left(  {\frac{\pi}{2\bar{K}}}\right)  ^{2}\mathrm{sech}%
^{2}\left[  \frac{\pi}{2\bar{K}}\left(  x-2K\ell\right)  \right]  , \label{PT}%
\end{equation}
where $\bar{K}=K(\bar{\kappa})$, $\bar{E}=E(\bar{\kappa})$ (the complete
elliptic integral of the second kind) with $\bar{\kappa}=\sqrt{1-\kappa^{2}}$
being a complementary modulus. A single particle traveling in this potential
has $n$ bound states and one perfectly transmitted (reflectionless)
\ scattering state\cite{Flugge}. When the potentials form a lattice, the bound
state overlaps and the energy band may be formed. Even after the band
formation, the $n$ gaps between the bound states and the scattering continuum
retain. Therefore, the resulting band is split into the lower valence bands
and the upper conduction band. The scenario for the case $n=1$ has been
discussed in the seminal work by Sutherland \cite{Sutherland1973}.

To obtain a closed analytic form of the Bloch wave solution in a
Hermite--Halphen form (we call this Hermite--Halphen--Bloch solution), we need
to specify the paths of the complex parameters $\alpha_{1},\alpha
_{2},...\alpha_{n}$ on a complex plane which satisfies the conditions
(\ref{LameSolution}), (\ref{Condition1}), (\ref{Condition2}),
(\ref{wavenumber}), and (\ref{condition_for_wave_number}). The case of $n=1$
has been well known \cite{Sutherland1973} for an arbitrary $0\leq\kappa\leq1$,
but the extension to $n>1$ is even numerically nontrivial. Fortunately, a
recent paper by Maier~\cite{Maier2008} offers a method alternative to the
Hermite--Halphen construction, by using the Hermite--Krishever Ansatz. Based
on this ansatz, Maier succeeded in obtaining the band dispersion relations for
any integer $n$ in terms of the $n=1$ relations. However, in viewpoint of
physical applications it may be still useful to seek a closed form of the
Hermite--Halphen--Bloch solution for $n>1$, which is still absent as far as
the authors know. In this paper, we report on how to construct the
Hermite--Halphen--Bloch solution for $n=2$.

Based on the Hermite--Krichever Ansatz, which expresses a solution of the
Lam\'{e} equation in terms of the known ${n=1}$ solution,
Maier\cite{Maier2008} derived the spectral polynomial,
\begin{eqnarray}
L_{2}(\varepsilon|\kappa)&=&\left[  \varepsilon^{2}-4(\kappa^{2}+1)\varepsilon
+12\kappa^{2}\right]  (\varepsilon-\kappa^{2}-1)\nonumber\\
&\times&(\varepsilon-4\kappa^{2}-1)(\varepsilon
-\kappa^{2}-4)=0,
\end{eqnarray}
whose roots
\begin{align}
\varepsilon_{1} &  =2\left(  \kappa^{2}+1\right)  -2\sqrt{\kappa^{4}%
-\kappa^{2}+1},\label{e0}\\
\varepsilon_{2} &  =\kappa^{2}+1,\label{e1}\\
\varepsilon_{3} &  =4\kappa^{2}+1,\label{e2}\\
\varepsilon_{4} &  =\kappa^{2}+4,\label{e3}\\
\varepsilon_{5} &  =2\left(  \kappa^{2}+1\right)  +2\sqrt{\kappa^{4}%
-\kappa^{2}+1},\label{e4}%
\end{align}
give the energy eigenvalues at the band edges, i.e., three allowed bands
consist of the first band $\varepsilon_{1}\leq\varepsilon\leq\varepsilon_{2}$
($\left\vert k\right\vert \leq\pi/2K$)$,$ the second band $\varepsilon_{3}%
\leq\varepsilon\leq\varepsilon_{4}$ ($\pi/2K\leq\left\vert k\right\vert
\leq\pi/K$)\ and the third one $\varepsilon_{5}\leq\varepsilon$ ($\pi
/K\leq\left\vert k\right\vert $). The construction of the bands and the band
edges values for the $n=2$ Lam\'{e} equation in the Weierstrass form is given
in the Appendix C.

Our goal is to determine the pathways of $(\alpha_{1},\alpha_{2})$ on a
complex plane which parametrize these three bands. Noting
\begin{align}
\operatorname{Re}\left[  Z(iy)\right]   &  =0,\label{Z_reality1}\\
\operatorname{Re}\left[  Z(K+iy)\right]   &  =0,\label{Z_reality2}\\
\operatorname{Re}\left[  Z(-x+iy)+Z(x+iy)\right]   &  =0, \label{Z_reality3}%
\end{align}
where $x,y\in\mathbb{R}$ and taking heed that the zeta function $Z(x)$ is a
singly periodic function with the period $2K$, we can locate the following trajectories.

\bigskip

\begin{figure}[h]
\begin{center}
\includegraphics[width=85mm]{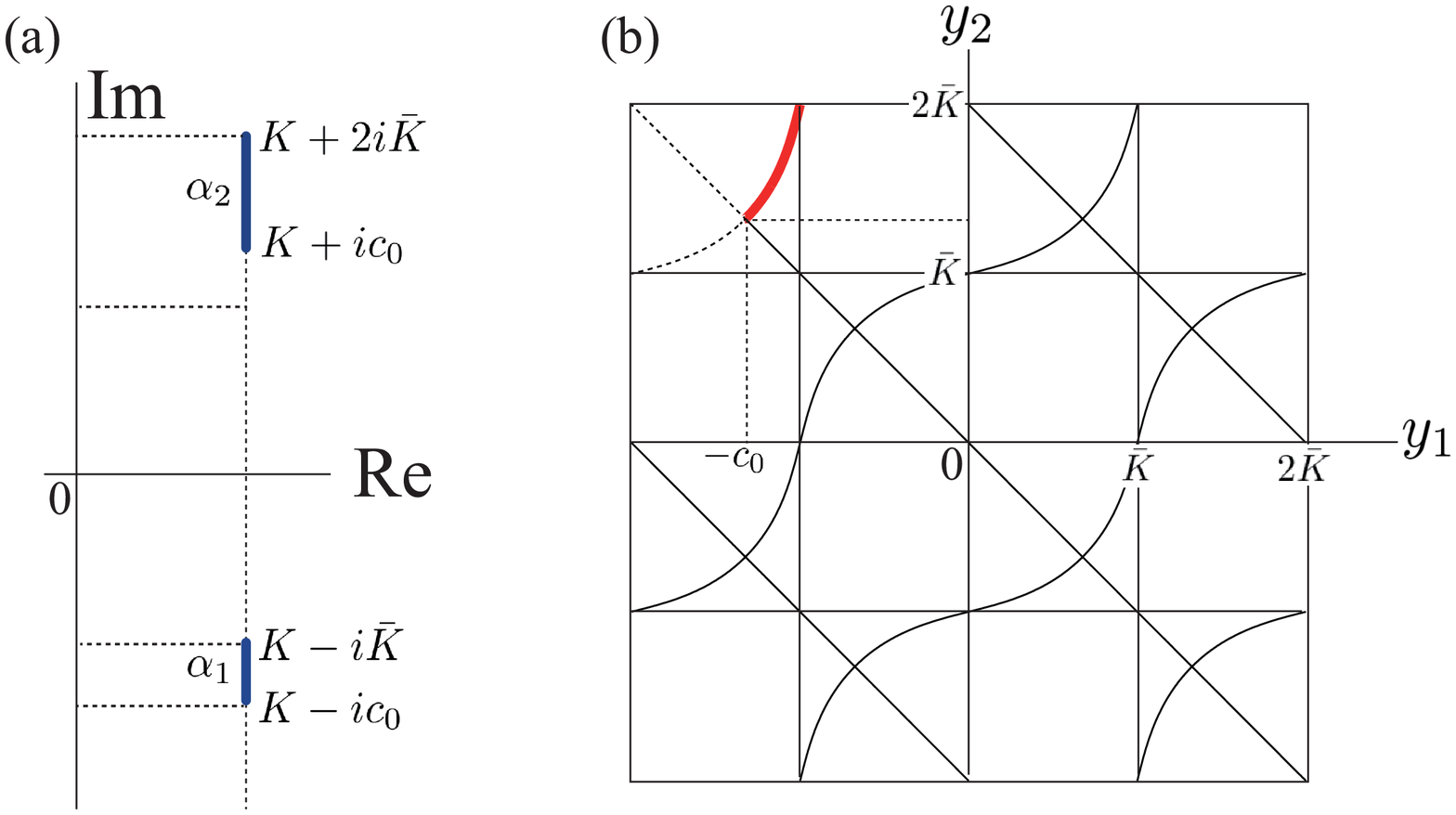}
\end{center}
\caption{(a) The pathways for $\alpha=K+iy_{1}$ and $\alpha_{2}=K+iy_{2}$. We
choose $\kappa^{2}=1/2$ to obtain these plot. (b) Trajectory of the point
$\left(  y_{1},y_{2}\right)  $. }%
\label{Path1}%
\end{figure}

\noindent(1) For the first band ($\varepsilon_{1}\leq\varepsilon
\leq\varepsilon_{2}$), we take%
\begin{equation}
\alpha_{1}=K+iy_{1},\alpha_{2}=K+iy_{2},
\end{equation}
where $(y_{1},y_{2})$ lie in a fundamental region ($-3\bar{K}/2<y_{1}\leq
-\bar{K}$, $\bar{K}<y_{2}\leq2\bar{K}$). The condition (\ref{Condition2})
becomes%
\begin{equation}
\frac{\overline{\text{$\mathrm{sn}$}}y_{1}\overline{\mathrm{c}%
\text{$\mathrm{n}$}}y_{1}}{\overline{\mathrm{d}\text{$\mathrm{n}$}}^{3}y_{1}%
}+\frac{\overline{\text{$\mathrm{sn}$}}y_{2}\overline{\mathrm{c}%
\text{$\mathrm{n}$}}y_{2}}{\overline{\mathrm{d}\text{$\mathrm{n}$}}^{3}y_{2}%
}=0, \label{Path_for_Band1}%
\end{equation}
where we used the notation $\overline{\text{$\mathrm{sn}$}}y_{1}%
=\mathrm{sn}\left(  y_{1},\bar{\kappa}\right)  $ and alike. In Fig.\ref{Path1}%
(a), we explicitly show the pathways for $(\alpha_{1}=K+iy_{1},\alpha
_{2}=K+iy_{2})$, and in Fig.\ref{Path1}(b) we show the corresponding
trajectory of $\left(  y_{1},y_{2}\right)  $. 
On this segment, the energy becomes%
\begin{align}
\varepsilon &  =2\kappa^{2}+\bar{\kappa}^{2}\frac{\overline{\text{$\mathrm{cn}%
$}}^{2}y_{1}}{\overline{\text{$\mathrm{dn}$}}^{2}y_{1}}+\bar{\kappa}^{2}%
\frac{\overline{\text{$\mathrm{cn}$}}^{2}y_{2}}{\overline{\text{$\mathrm{dn}$%
}}^{2}y_{2}}+2\bar{\kappa}^{4}\frac{\overline{\text{$\mathrm{sn}$}}%
y_{1}\overline{\text{$\mathrm{cn}$}}y_{1}\overline{\text{$\mathrm{sn}$}}%
y_{2}\overline{\text{$\mathrm{cn}$}}y_{2}}{\overline{\text{$\mathrm{dn}$}%
}y_{1}\overline{\text{$\mathrm{dn}$}}y_{2}}\nonumber\\
&  =2\kappa^{2}+\bar{\kappa}^{2}\frac{\overline{\text{$\mathrm{cn}$}}^{2}%
y_{1}}{\overline{\text{$\mathrm{dn}$}}^{2}y_{1}}+\bar{\kappa}^{2}%
\frac{\overline{\text{$\mathrm{cn}$}}^{2}y_{2}}{\overline{\text{$\mathrm{dn}$%
}}^{2}y_{2}}-2\bar{\kappa}^{4}\frac{\overline{\text{$\mathrm{sn}$}}^{2}%
y_{1}\overline{\text{$\mathrm{cn}$}}^{2}y_{1}}{\overline{\text{$\mathrm{dn}$}%
}^{4}y_{1}}\overline{\text{$\mathrm{dn}$}}^{2}y_{2}, \label{enegy1}%
\end{align}
where we used the condition (\ref{Path_for_Band1}).

The band bottom corresponds to
\begin{equation}
\alpha_{1}=K-ic_{0},\text{ \ }\alpha_{2}=K+ic_{0},
\end{equation}
which gives
\begin{eqnarray}
k&=&-iZ(K-ic_{0})-iZ(K+ic_{0})+\frac{\pi}{K}\nonumber\\
&=&\frac{\pi}{K} =0, \qquad\left(
\text{mod} \quad\pi/K \right)
\end{eqnarray}
where we used%
\begin{eqnarray}
&&Z\left(  K+iy\right) \nonumber\\
&=&i\left(  \overline{\text{$\mathrm{dn}$}}y\overline
{\text{$\mathrm{sc}$}}y-\bar{Z}(y)-\frac{\pi}{2K\bar{K}}y-\kappa^{2}
\frac{\overline{\text{$\mathrm{sn}$}}y}{\overline{\text{$\mathrm{cn}$}}y
\overline{\text{$\mathrm{dn}$}}y} \right)  .
\end{eqnarray}
Here, $\bar{Z}=Z(\bar{\kappa})$ and $c_{0}$ is determined by (\ref{e0}) and
(\ref{enegy1}), i.e.,%
\begin{equation}
\varepsilon_{1}=2\kappa^{2}+2\bar{\kappa}^{2}\frac{\overline
{\text{$\mathrm{cn}$}}^{2}c_{0}}{\overline{\text{$\mathrm{dn}$}}^{2}c_{0}%
}-2\bar{\kappa}^{4}\frac{\overline{\text{$\mathrm{sn}$}}^{2}c_{0}%
\overline{\text{$\mathrm{cn}$}}^{2}c_{0}}{\overline{\text{$\mathrm{dn}$}}%
^{2}c_{0}}=2\kappa^{2}+2\bar{\kappa}^{2}\overline{\text{$\mathrm{cn}$}}%
^{2}c_{0}, \label{epsilon0co}%
\end{equation}
that gives the location
\begin{eqnarray}
\bar{\kappa}^{2}\overline{\text{$\mathrm{cn}$}}^{2}c_{0}
=1-\sqrt{\kappa
^{4}-\kappa^{2}+1}.
\end{eqnarray}
The band top $\varepsilon_{2}$\ is given by $\alpha_{1}=K-i\bar{K},$
\ $\alpha_{2}=K+2i\bar{K}$, which actually gives $k=\pi/2K,$ $\varepsilon
=\varepsilon_{2}.$

\bigskip

\noindent(2) For the second band ($\varepsilon_{3}\leq\varepsilon
\leq\varepsilon_{4}$), we take%
\begin{equation}
\alpha_{1}=K+iy_{1},\ \ \alpha_{2}=iy_{2},
\end{equation}
where $(y_{1},y_{2})$ lie in the fundamental region ($0\leq y_{1}\leq\bar{K}$,
$0\leq y_{2}\leq\bar{K}/2$). The condition (\ref{Condition2}) becomes%
\begin{equation}
\bar{\kappa}^{2}\frac{\overline{\text{$\mathrm{sn}$}}y_{1}\overline
{\mathrm{c}\text{$\mathrm{n}$}}y_{1}}{\overline{\mathrm{d}\text{$\mathrm{n}$}%
}^{3}y_{1}}=\frac{\overline{\mathrm{s}\text{$\mathrm{n}$}}y_{2}\overline
{\mathrm{d}\text{$\mathrm{n}$}}y_{2}}{\overline{\mathrm{c}\text{$\mathrm{n}$}%
}^{3}y_{2}}, \label{Path_for_Band2}%
\end{equation}
On this segment, the energy becomes%
\begin{equation}
\varepsilon=2\kappa^{2}+\bar{\kappa}^{2}\frac{\overline{\mathrm{c}%
\text{$\mathrm{n}$}}^{2}y_{1}}{\overline{\mathrm{d}\text{$\mathrm{n}$}}%
^{2}y_{1}}+\frac{\overline{\mathrm{d}\text{$\mathrm{n}$}}^{2}y_{2}}%
{\overline{\mathrm{c}\text{$\mathrm{n}$}}^{2}y_{2}}+2\bar{\kappa}^{2}%
\frac{\overline{\text{$\mathrm{sn}$}}y_{1}\overline{\text{$\mathrm{cn}$}}%
y_{1}\overline{\mathrm{d}\text{$\mathrm{n}$}}y_{2}}{\overline{\mathrm{s}%
\text{$\mathrm{n}$}}y_{2}\overline{\mathrm{c}\text{$\mathrm{n}$}}%
y_{2}\overline{\text{$\mathrm{dn}$}}y_{1}},
\end{equation}

\begin{figure}[h]
\begin{center}
\includegraphics[width=85mm]{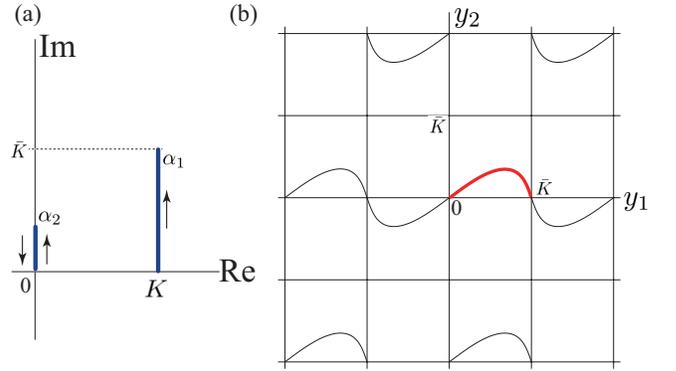}
\end{center}
\caption{(a) The pathways for $\alpha_{1}=K+iy_{1}$ and $\alpha_{2}=iy_{2}$.
We choose $\kappa^{2}=1/2$ to obtain these plot. (b) Trajectory of the point
$\left(  y_{1},y_{2}\right)  $. }%
\label{Path2}%
\end{figure}Noting (\ref{Path_for_Band2}), we have%

\begin{align}
\varepsilon=2\kappa^{2}+\bar{\kappa}^{2}\frac{\overline{\mathrm{c}%
\text{$\mathrm{n}$}}^{2}y_{1}}{\overline{\mathrm{d}\text{$\mathrm{n}$}}%
^{2}y_{1}}+\frac{\overline{\mathrm{d}\text{$\mathrm{n}$}}^{2}y_{2}}%
{\overline{\mathrm{c}\text{$\mathrm{n}$}}^{2}y_{2}}+2\overline{\mathrm{d}%
\text{$\mathrm{n}$}}^{2}y_{1}\frac{\overline{\mathrm{d}\text{$\mathrm{n}$}%
}^{2}y_{2}}{\overline{\mathrm{c}\text{$\mathrm{n}$}}^{4}y_{2}} \label{ene2}%
\end{align}
In Fig.\ref{Path2}(a), we show the pathways for $(\alpha_{1}=K+iy_{1}%
,\ \ \alpha_{2}=iy_{2})$, and in Fig.\ref{Path2}(b) we show the corresponding
trajectory of $\left(  y_{1},y_{2}\right)  $. The band bottom ($k=\pi/2K$)
corresponds to $y_{1}=\bar{K},$ \ \ \ \ $y_{2}=0$ while the band top
($k=\pi/K$) does to $y_{1}=0,$ $y_{2}=0.$

\bigskip

\begin{figure}[h]
\begin{center}
\includegraphics[width=45mm]{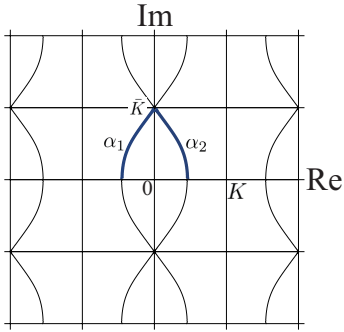}
\end{center}
\caption{Relevant branches of $\alpha_{1}=-x+iy$ and $\text{ \ }\alpha
_{2}=x+iy$ which properly reproduces the third band. The case $\kappa^{2}=1/2$
is presented.}%
\label{Path3}%
\end{figure}

\noindent(3) For the third band ($\varepsilon_{5}\leq\varepsilon$), we take%
\begin{equation}
\alpha_{1}=-x+iy,\text{ \ }\alpha_{2}=x+iy,
\end{equation}
where $(y_{1},y_{2})$ lie in the fundamental region ($0\leq x\leq K/2$, $0\leq
y\leq\bar{K}$). The condition (\ref{Condition2}) becomes%
\begin{equation}
\overline{\text{\textrm{sn}}}^{2}y=\dfrac{\kappa^{2}\text{\textrm{sn}}%
^{2}x\text{\textrm{cn}}^{2}x+\text{\textrm{sn}}^{2}x\text{\textrm{dn}}%
^{2}x-\text{\textrm{cn}}^{2}x\text{\textrm{dn}}^{2}x}{\left(  {\bar{\kappa}}
^{2} \text{\textrm{sn}}^{2}x-\text{\textrm{cn}}^{2}x - \kappa^{2}%
\text{\textrm{sn}}^{2}x \text{\textrm{cn}}^{2}x \right)  \text{\textrm{dn}%
}^{2}x}.
\end{equation}
In Fig. \ref{Path3}, we show the pathways for $(\alpha_{1}=-x+iy,$
\ $\alpha_{2}=x+iy)$. On this segment,
\begin{align}
\varepsilon=2\kappa^{2}+\mathrm{dn}^{2}\alpha_{1}+\mathrm{dn}^{2}\alpha
_{2}-2\frac{\text{$\mathrm{cn}$}\alpha_{1}\,\text{$\mathrm{dn}$}\alpha
_{1}\text{$\mathrm{cn}$}\alpha_{2}\,\text{$\mathrm{dn}$}\alpha_{2}%
}{\text{$\mathrm{sn}$}\alpha_{1}\text{$\mathrm{sn}$}\alpha_{2}} \label{ene3}%
\end{align}

The band bottom ($k=\pi/K$) corresponds to%
\begin{equation}
\alpha_{1}=-x_{0},\text{ \ }\alpha_{2}=x_{0},
\end{equation}
where $x_{0}$ is determined by (\ref{ene3}) and (\ref{e4}), i.e.,%
\begin{equation}
\text{$\mathrm{dn}$}^{2}x_{0}+\frac{\text{$\mathrm{cn}$}^{2}x_{0}%
\text{$\mathrm{dn}$}^{2}x_{0}}{\text{$\mathrm{sn}$}^{2}x_{0}}=1+\sqrt
{\kappa^{4}-\kappa^{2}+1}%
\end{equation}
and the band top ($k\rightarrow\infty$) \ is given by $\alpha_{1}=\alpha
_{2}=i\bar{K}$, because $Z(\alpha)$ and $\mathrm{dn}(\alpha)$ have a pole at
$\alpha=i\bar{K}$.

\begin{figure}[h]
\begin{center}
\includegraphics[width=80mm]{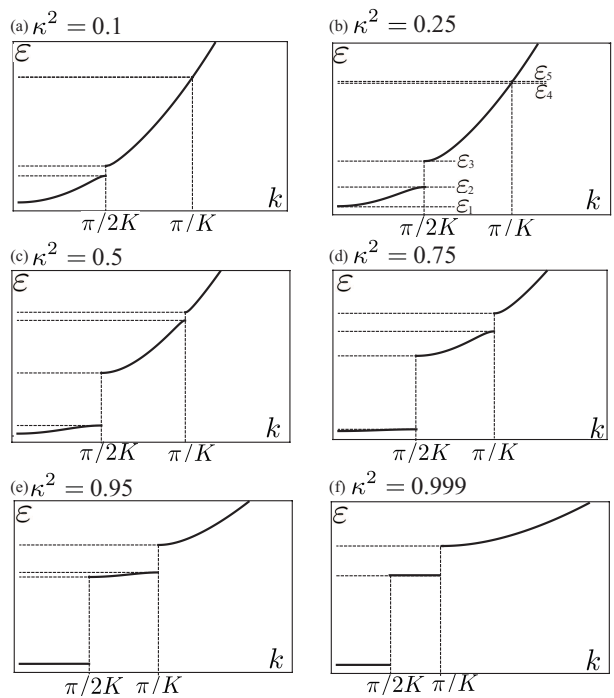}
\end{center}
\caption{The Bloch band dispersions for various values of $\kappa$.}%
\label{Band}%
\end{figure}

Using the obtained results, we enable to compute the Bloch band dispersion for
an arbitrary $0\leq\kappa\leq1$. In Fig. \ref{Band}, we show the results. It
is clearly seen that as the $\kappa$ approaches unity the lower two bands
become flatter and finally utterly flat at the limit of $\kappa\rightarrow1$.
This phenomena is easily understood as follows. As $\kappa$ increases from $0$
to $1$, the period of the Lam\'{e} potential $2K$ increases from $\pi$ to
$\infty$. Therefore, the overlap between the modified P\"{o}schl-Teller
potentials [see Eq. (\ref{PT})] becomes smaller. Consequently, two bound
states originating from an independent modified P\"{o}schl-Teller potential
with $n=2$ become more localized. On the other hand, for smaller $\kappa$, the
overlap becomes large and the bound states form energy bands with a larger
band width. The connection between these results and that of obtained in Ref.
\cite{Baryakhtar2006} in terms of Weierstrass elliptic functions is discussed
in Appendix D.

In summary, in this paper, we succeeded in constructing the
Hermite--Halphen--Bloch solution for $n=2$ Lam\'e equation and obtained closed
formulae which give the band dispersion relation. From a physical viewpoint,
these dispersions give fluctuation spectra around the soliton lattice solution
of the classical $\phi^{4}$ -field theory. We hope our results may be useful
to promote physical analysis related with this model.

\begin{acknowledgments}
This work was supported by the Government of Russian Federation Program
02.A03.21.0006 and by Japan Society for the Promotion of Science Grants-in-Aid
for Scientific Research (S) Nos. 25220803. We are also supported by Center for
Chiral Science, Hiroshima University.
\end{acknowledgments}

\appendix

\section{The momentum in the Weierstrass form of the Lam\'{e} equation}

Henceforth, we use the notations of the paper \cite{Baryakhtar2006}, where the
potential $U(x) = -n(n+1)
\raisebox{.15\baselineskip}{\Large\ensuremath{\wp}}(ix+\omega)$, determined
via the Weierstrass elliptic function
$\raisebox{.15\baselineskip}{\Large\ensuremath{\wp}}$, has the period $2
|\omega^{^{\prime}}|$, the real and the imaginary half periods are $\omega$
and $\omega^{\prime}$, respectively.

The fundamental solution of the Lam\'e equation with $n=2$ in the Weierstrass
form is given by (see Ref.~\cite{Whittaker-Watson})
\begin{eqnarray}
\Lambda_{1} (ix+\omega) &=& \prod_{r=1}^{2} \left\{  \frac{\sigma(t_{r}%
+ix+\omega)}{\sigma(ix+\omega) \sigma(t_{r})} \right\}\nonumber\\
&\times&  \exp\left\{
-(ix+\omega) \sum_{r=1}^{2} \zeta(t_{r}) \right\}  ,
\end{eqnarray}
where $\sigma$ and $\zeta$ are the Weierstrass's sigma and zeta functions, respectively.

The factor $\exp{ \left\{  -\omega\sum_{r=1}^{2} \zeta(t_{r})\right\}  }$ is
the constant and it can be dropped that gives
\begin{eqnarray}
\Lambda_{1} (ix+\omega) &=& \prod_{r=1}^{2} \left\{  \frac{\sigma(t_{r}%
+ix+\omega)}{\sigma(ix+\omega) \sigma(t_{r})} \right\}
 \nonumber\\
&\times&  \exp\left\{  -ix
\sum_{r=1}^{2} \zeta(t_{r}) \right\}  .
\end{eqnarray}

To convert the solution into the Bloch form $\Lambda_{1} (ix+\omega) = u(x)
e^{-ikx}$, where the function $u(x)$ has the period of the potential $u(x+2
|\omega^{^{\prime}}|) = u(x-2i\omega^{^{\prime}}) = u(x)$, the constants
$\alpha_{r}$ are introduced
\begin{eqnarray}
\Lambda_{1} (ix+\omega) &=& \prod_{r=1}^{2} \left\{  \frac{\sigma(t_{r}%
+ix+\omega)}{\sigma(ix+\omega) \sigma(t_{r})} e^{-ix \alpha_{r}} \right\}
\nonumber\\
&\times&\exp\left\{  -ix \sum_{r=1}^{2} \left(  \zeta(t_{r}) -\alpha_{r} \right)
\right\} \nonumber\\
&=& u_{r}(x) \exp\left\{  -ikx \right\}  .
\end{eqnarray}

By using the property
\begin{equation}
\sigma(x+2\omega^{^{\prime}}) = - e^{2\eta_{2} (x+\omega^{^{\prime}})}
\sigma(x),
\end{equation}
where the constant $\eta_{2} = \zeta(\omega^{^{\prime}})$, we obtain the
periodicity $u_{r}(x- 2i\omega^{^{\prime}}) = u_{r}(x)$ provided
\begin{equation}
\alpha_{r} = \frac{\eta_{2}}{\omega^{^{\prime}}} t_{r}.
\end{equation}

Thereby, the momentum is given by
\begin{equation}
k=\sum_{r=1}^{2}\left(  \zeta(t_{r})-\alpha_{r}\right)  =\sum_{r=1}^{2}\left(
\zeta(t_{r})-\frac{\eta_{2}}{\omega^{^{\prime}}}t_{r}\right)  .
\end{equation}

\section{Derivation of Eq.(\ref{decompositionfourmula})}

\begin{figure}[h]
\begin{center}
\includegraphics[width=65mm]{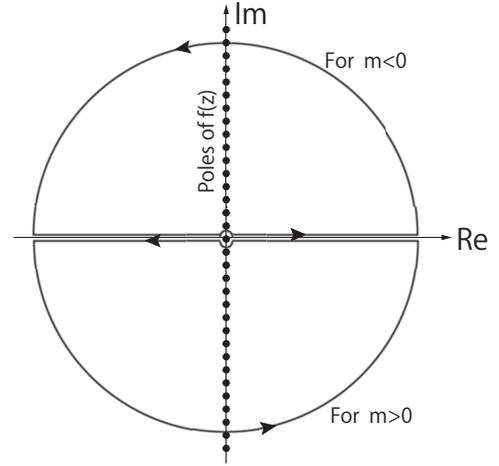}
\end{center}
\caption{Contours in the complex plane to compute the integral (\ref{integral}).}%
\label{Path}%
\end{figure}

We start with the Fourier series for the Zeta function\cite{Whittaker-Watson}%
,
\begin{equation}
Z(x)=\frac{\pi}{K}{\displaystyle\sum_{n=1}^{\infty}}\frac{\sin(n\pi
x/K)}{\sinh(n\pi\bar{K}/K)},
\end{equation}
and obtain
\begin{eqnarray}
\kappa^{2}\mathrm{\mathrm{sn}}^{2}x&=&1-\mathrm{\mathrm{dn}}^{2}x=1-\left(
\frac{E}{K}+\frac{dZ(x)}{dx}\right)  \nonumber\\
&=&1-\frac{E}{K}-\left(  \frac{\pi}%
{K}\right)  ^{2}{\displaystyle\sum_{n=1}^{\infty}}f(x,n),
\end{eqnarray}
where%
\begin{equation}
f(x,n)=\frac{n\cos(n\pi x/K)}{\sinh(n\pi\bar{K}/K)}.
\end{equation}
Noting $\lim_{n\rightarrow0}\left[  n\cos(n\pi x/K)/\sinh(n\pi\bar
{K}/K)\right]  =K/(\pi\bar{K})$ and the Legendere's identity $K\bar{E}+\bar
{K}E-K\bar{K}=\pi/2$, we have
\begin{equation}
\kappa^{2}\mathrm{\mathrm{sn}}^{2}x=\frac{\bar{E}}{\bar{K}}-\frac{\pi^{2}%
}{2K^{2}}%
{\displaystyle\sum_{n=-\infty}^{\infty}}
f(x,n) \label{intermediate}%
\end{equation}
The 2nd term on the r.h.s. is computed by using the Poisson summation formula,%
\begin{equation}
S(x)\equiv%
{\displaystyle\sum_{n=-\infty}^{\infty}}
f(x,n)=%
{\displaystyle\sum_{m=-\infty}^{\infty}}
\int_{-\infty}^{\infty}f(x,\zeta)e^{-2\pi im\zeta}d\zeta.
\end{equation}
The integral is evaluated as%
\begin{equation}
I(x)=\int_{-\infty}^{\infty}f(x,\zeta)e^{-2\pi im\zeta}d\zeta=%
{\displaystyle\oint_{C}}
f(x,z)e^{-2\pi imz}dz,\label{integral}
\end{equation}
where $C$ is a upper and lower semicircle on the complex $z$-plane for $m<0$
and $m>0$, respectively [see Fig. \ref{Path}]. 

Picking up residues at the poles of $f(x,z)$,
$z_{\ell}=\left(  iK/\bar{K}\right)  \ell$, we obtain%
\begin{eqnarray}
&&{\displaystyle\oint_{C}}f(x,z)e^{-2\pi imz}dz
\nonumber\\
&=&\frac{1}{2}\left(  \frac{K}%
{\bar{K}}\right)  ^{2}{\displaystyle\sum_{\ell=-\infty}^{\infty}}%
\text{sech}^{2}\left[  \frac{\pi}{2\bar{K}}\left(  x-2K\ell\right)  \right]  .
\end{eqnarray}
Plugging this result into Eq. (\ref{intermediate}), we arrive at Eq.
(\ref{decompositionfourmula}).

\section{The spectrum of the Lam\'e equation in the Weierstrass form}

The Lam\'e equation in the Weierstrass form
\begin{equation}
\label{Lame}\frac{d^{2} \Lambda}{d u^{2}} = \left\{  n(n+1)
\raisebox{.15\baselineskip}{\Large\ensuremath{\wp}}(u) + B \right\}  \Lambda
\end{equation}
has two independent solutions
\begin{equation}
\label{SolWLam}\Lambda_{1,2} = \sqrt{X} \exp\left\{  \mp\mathcal{Q} \int
\frac{du}{X} \right\}  ,
\end{equation}
where $\mathcal{Q}$ is the Wronskian, and $X$ is the product of the pair, $X =
\Lambda_{1} \Lambda_{2}$, obeying the equation \cite{Whittaker-Watson}
\begin{equation}
\label{eqforX}\frac{d^{3} X}{du^{3}} - 4 \left\{  n(n+1)
\raisebox{.15\baselineskip}{\Large\ensuremath{\wp}}(u) + B \right\}  \frac
{dX}{du} - 2 n(n+1)
\raisebox{.15\baselineskip}{\Large\ensuremath{\wp}}^{^{\prime}}(u) X =0.
\end{equation}

For the case $n=2$ the form of $X$ can be taken as
\begin{equation}
\label{ansatz}X(u) = C_{0}
\raisebox{.15\baselineskip}{\Large\ensuremath{\wp}}^{2} (u) + C_{1}
\raisebox{.15\baselineskip}{\Large\ensuremath{\wp}}(u) + C_{2},
\end{equation}
where $C_{i}$ are the constants, that corresponds to the solution in
descending powers of $\raisebox{.15\baselineskip}{\Large\ensuremath{\wp}}(u) -
e_{2}$ (see Ref.~\cite{Whittaker-Watson}).

It transforms (\ref{eqforX}) into
\begin{equation}
\label{toFindX}- \partial^{3}_{x} X + 4 \left\{  6
\raisebox{.15\baselineskip}{\Large\ensuremath{\wp}}(ix +\omega) + E \right\}
\partial_{x} X + 12 \partial_{x}
\raisebox{.15\baselineskip}{\Large\ensuremath{\wp}} (ix +\omega) X =0,
\end{equation}
where the notations of Ref. \cite{Baryakhtar2006} are adopted, $u = ix
+\omega$, $B = - E$ and $\raisebox{.15\baselineskip}{\Large\ensuremath{\wp}}
(u) = - \raisebox{.15\baselineskip}{\Large\ensuremath{\wp}} (ix+\omega)$,
where $E$ is the energy.

By calculating with the aid of Eq. (\ref{ansatz})
\begin{equation}
\label{1stD}\partial_{x} X = 2C_{0}
\raisebox{.15\baselineskip}{\Large\ensuremath{\wp}} \partial_{x}
\raisebox{.15\baselineskip}{\Large\ensuremath{\wp}} + C_{1} \partial_{x}
\raisebox{.15\baselineskip}{\Large\ensuremath{\wp}},
\end{equation}
\begin{equation}
\label{3rdD}\partial^{3}_{x} X = 6 C_{0} \partial_{x}
\raisebox{.15\baselineskip}{\Large\ensuremath{\wp}} \partial^{2}_{x}
\raisebox{.15\baselineskip}{\Large\ensuremath{\wp}} + 2C_{0}
\raisebox{.15\baselineskip}{\Large\ensuremath{\wp}} \partial^{3}_{x}
\raisebox{.15\baselineskip}{\Large\ensuremath{\wp}} + C_{1} \partial^{3}_{x}
\raisebox{.15\baselineskip}{\Large\ensuremath{\wp}},
\end{equation}
and excluding the higher-order derivatives through the identities
\cite{Whittaker-Watson}
\begin{equation}
\partial^{2}_{x} \raisebox{.15\baselineskip}{\Large\ensuremath{\wp}} = 6
\raisebox{.15\baselineskip}{\Large\ensuremath{\wp}}^{2} - \frac12 g_{2},
\qquad\partial^{3}_{x} \raisebox{.15\baselineskip}{\Large\ensuremath{\wp}} =
12 \raisebox{.15\baselineskip}{\Large\ensuremath{\wp}} \partial_{x}
\raisebox{.15\baselineskip}{\Large\ensuremath{\wp}},
\end{equation}
where $g_{2}$ is the invariant, we obtain from Eq. (\ref{toFindX})
\begin{eqnarray}
&&- \left[  6C_{0} \raisebox{.15\baselineskip}{\Large\ensuremath{\wp}}^{\prime
}\left(  6 \raisebox{.15\baselineskip}{\Large\ensuremath{\wp}}^{2} - \frac12
g_{2} \right)  + 24 C_{0}
\raisebox{.15\baselineskip}{\Large\ensuremath{\wp}}^{2}
\raisebox{.15\baselineskip}{\Large\ensuremath{\wp}}^{\prime}+ 12 C_{1}
\raisebox{.15\baselineskip}{\Large\ensuremath{\wp}}
\raisebox{.15\baselineskip}{\Large\ensuremath{\wp}}^{\prime}\right] 
\nonumber\\ 
&+& 4
\left(  6\raisebox{.15\baselineskip}{\Large\ensuremath{\wp}} + E \right)
\left(  2C_{0} \raisebox{.15\baselineskip}{\Large\ensuremath{\wp}}
\raisebox{.15\baselineskip}{\Large\ensuremath{\wp}}^{\prime}+ C_{1}
\raisebox{.15\baselineskip}{\Large\ensuremath{\wp}}^{\prime}\right)
\end{eqnarray}
\begin{equation}
+ 12 \raisebox{.15\baselineskip}{\Large\ensuremath{\wp}}^{\prime}\left(  C_{0}
\raisebox{.15\baselineskip}{\Large\ensuremath{\wp}}^{2} + C_{1}
\raisebox{.15\baselineskip}{\Large\ensuremath{\wp}} + C_{2} \right)  = 0.
\end{equation}
By finding the coefficients of the same powers of
$\raisebox{.15\baselineskip}{\Large\ensuremath{\wp}}$ and
$\raisebox{.15\baselineskip}{\Large\ensuremath{\wp}}^{\prime}$ we get the
relations $C_{1} ~=~ - \left(  E/3\right)  C_{0}$ and $C_{2} = - \left(
E/3\right)  C_{1} - \left(  C_{0}/4\right)  g_{2}$.

The choice $C_{0}=18$ of the Ref. \cite{Baryakhtar2006} results in
\begin{equation}
\label{XonE}X = 18 \raisebox{.15\baselineskip}{\Large\ensuremath{\wp}}^{2}
(ix+\omega) - 6 E \raisebox{.15\baselineskip}{\Large\ensuremath{\wp}}
(ix+\omega) + 2 E^{2} - \frac92 g_{2}.
\end{equation}
The polynomial can be presented as
\begin{equation}
X = 18 \left(  \raisebox{.15\baselineskip}{\Large\ensuremath{\wp}} (ix+\omega)
- \raisebox{.15\baselineskip}{\Large\ensuremath{\wp}}(t_{1}) \right)  \left(
\raisebox{.15\baselineskip}{\Large\ensuremath{\wp}} (ix+\omega) -
\raisebox{.15\baselineskip}{\Large\ensuremath{\wp}}(t_{2}) \right)  ,
\end{equation}
where
\begin{equation}
\raisebox{.15\baselineskip}{\Large\ensuremath{\wp}}(t_{1}) = \frac{E}{6} +
\frac{1}{2\sqrt{3}} \sqrt{3 g_{2} - E^{2}},
\end{equation}
\begin{equation}
\raisebox{.15\baselineskip}{\Large\ensuremath{\wp}}(t_{2}) = \frac{E}{6} -
\frac{1}{2\sqrt{3}} \sqrt{3 g_{2} - E^{2}}%
\end{equation}
give the parametric form for the spectrum of the Lam\'e equation with $n=2$
(see Ref.~\cite{Baryakhtar2006}).

To find the band edges we note that the phase factor in the solution
(\ref{SolWLam}) turns into zero at the band edges, what is equivalent to the
requirement $\mathcal{Q}=0$.

Given $X$, the Wronskian can be found through the relation
\cite{Whittaker-Watson}
\begin{equation}
\label{Wron}n(n+1) \raisebox{.15\baselineskip}{\Large\ensuremath{\wp}}(u) + B
= \frac{1}{2X} \frac{d^{2} X}{d u^{2}} - \frac{1}{4X^{2}} \left(  \frac{d X}{d
u} \right)  ^{2} + \frac{\mathcal{Q}^{2}}{X^{2}},
\end{equation}
which takes the form
\begin{equation}
\mathcal{Q}^{2} = - 4 \left\{  6
\raisebox{.15\baselineskip}{\Large\ensuremath{\wp}}(ix+\omega) +E \right\}
X^{2} + 2 X \partial^{2}_{x} X - \left(  \partial_{x} X \right)  ^{2}.
\end{equation}
in the notions of Ref. \cite{Baryakhtar2006}.

By using the result (\ref{XonE}) we obtain after simplification
\begin{eqnarray}
\mathcal{Q}^{2} = - \left(  E^{2} -3 g_{2} \right)  \left(  16 E^{3} -36 E
g_{2} + 108 g_{3} \right)
\end{eqnarray}
\begin{equation}
= - 16 (E- \sqrt{3 g_{2}}) (E+ \sqrt{3 g_{2}}) (E+3e_{1}) (E+3e_{2})
(E+3e_{3}),
\end{equation}
where we use the identities $e_{1} e_{2} + e_{2} e_{3} + e_{1} e_{3} = -
g_{2}/4$ and $e_{1} e_{2} e_{3} = g_{3}/4$. Here, $g_{3}$ is the invariant.

This yields the band edges
\begin{eqnarray}
&&\label{EdgesWeier}E_{1} = - \sqrt{3 g_{2}}, \quad E_{2} = -3 e_{1}, \nonumber\\
&& E_{3} = -3 e_{2}, \quad E_{4} = -3 e_{3}, \quad E_{5} = \sqrt{3 g_{2}}%
\end{eqnarray}
as given in Ref. \cite{Baryakhtar2006}

\section{The connection between Weierstrass's and Jacobi's forms of the
solutions}

The relationship between the results (\ref{e0},\ref{e1},\ref{e2}%
,\ref{e3},\ref{e4}) and (\ref{EdgesWeier}) is reached via the formula
\cite{Whittaker-Watson}
\begin{eqnarray}
\raisebox{.15\baselineskip}{\Large\ensuremath{\wp}} (u) = e_{3} + (e_{1} -
e_{3}) \text{ns}^{2} \left(  u \sqrt{e_{1} -e_{3}} \right)  .
\end{eqnarray}
The Jacobi's elliptic function having its modulus given by the equation
\begin{equation}
\label{kappa2}\kappa^{2} = \frac{e_{2}-e_{3}}{e_{1}-e_{3}}.
\end{equation}
The semiperiods $\omega_{1,2}$ of the Weierstrass functions are related with
$K$ and $\bar{K}$ by
\begin{equation}
\label{w1w2}\omega_{1} = \frac{K}{\sqrt{e_{1} -e_{3}}}, \qquad\omega_{2} = i
\frac{\bar{K}}{\sqrt{e_{1} -e_{3}}}.
\end{equation}

Let us compare, for instance, the results for the spectrum with $\kappa
^{2}=1/2$, when $K=\bar{K}$.

Given $\omega_{1} =K$, we get from Eq. (\ref{w1w2}) $e_{1} -e_{3}=1$. Then, as
follow from Eq. (\ref{kappa2}), $e_{2}-e_{3}=\kappa^{2}=1/2$.

By using the identity $e_{1} + e_{2} + e_{3} = 0$, we find
\begin{eqnarray}
e_{1} &=& \frac13 \left(  2 - \kappa^{2} \right)  = \frac12,\\
e_{2} &=& - \frac13 \left(  1 - 2 \kappa^{2} \right)  = 0,\\
e_{3} &=& - \frac13 \left(  1 + \kappa^{2} \right)  = - \frac12.
\end{eqnarray}
Therefore, the invariant
 $g_{2}= - 4 \left(  e_{1} e_{2} + e_{1} e_{3} + e_{2}
e_{3} \right)  = 1$.

According to Ref. \cite{Baryakhtar2006}, the width of the first band is [see
Eq.(\ref{EdgesWeier}) in Appendix B]
\begin{equation}
\sqrt{3g_{2}} -3 e_{1} \approx0.23
\end{equation}
that coincides with the result followed from Eqs. (\ref{e0},\ref{e1})
\begin{equation}
2 \sqrt{\kappa^{4} -\kappa^{2}+1} - (\kappa^{2}+1) \approx0.23.
\end{equation}

The value of the first gap in the Weierstrass form is given by $3 (e_{1} -
e_{2}) = 1.5$ that agrees with the Jacobian's result $3 \kappa^{2} = 1.5$, See
Eqs.(\ref{e1},\ref{e2}).

By similar way, we determine the width of the second band in the Weierstrass
form, $3(e_{2}-e_{3})=1.5$, and get the same result in the Jacobi's form, $3(1
- \kappa^{2}) = 1.5$, as predicted by Eqs. (\ref{e2},\ref{e3}).

At last, in the Weierstrass form the second gap equals, $\sqrt{3g_{2}} + 3
e_{3} \approx0.23$, that is in an utter accordance with the result of Eqs.
(\ref{e3},\ref{e4}), $\kappa^{2}-2 + 2 \sqrt{\kappa^{4} -\kappa^{2}+1}
\approx0.23$.
\end{document}